\documentclass[superscriptaddress,aps,twocolumn]{revtex4-1}
\usepackage{bm}
\usepackage{float}
\usepackage{amssymb}
\usepackage{amsmath}
\usepackage{multirow}
\usepackage{lineno}
\usepackage{graphicx}
\usepackage{comment}
\usepackage{soul}
\usepackage[colorlinks=true,citecolor=blue,linkcolor=blue]{hyperref}
\usepackage[usenames]{color}
\usepackage{subfigure}

\begin{document}

\title{High-Harmonic Generation and Optical Torque Interaction via Relativistic Diffraction of a Spatiotemporal Vortex Light}

\author{Ke Hu}
\affiliation{Tsung-Dao Lee Institute, Shanghai Jiao Tong University, Shanghai 201210, China}
%\affiliation{Collaborative Innovation Center of IFSA (CICIFSA), Shanghai Jiao Tong University, Shanghai 200240, China}
\affiliation{School of Physics, Hangzhou Normal University, Hangzhou 311121, China}
\author{Xinju Guo}
\affiliation{Tsung-Dao Lee Institute, Shanghai Jiao Tong University, Shanghai 201210, China}
\affiliation{Collaborative Innovation Center of IFSA (CICIFSA), Shanghai Jiao Tong University, Shanghai 200240, China}
\author{Longqing Yi}
\thanks{lqyi@sjtu.edu.cn}
\affiliation{Tsung-Dao Lee Institute, Shanghai Jiao Tong University, Shanghai 201210, China}
\affiliation{Collaborative Innovation Center of IFSA (CICIFSA), Shanghai Jiao Tong University, Shanghai 200240, China}
\affiliation{Key Laboratory for Laser Plasmas (Ministry of Education), School of Physics and Astronomy, Shanghai Jiao Tong University, Shanghai 200240, China}

\date{\today}

\begin{abstract}

Diffraction of a relativistically-strong light can produce high-order harmonics via the ``relativistic oscillating window" mechanism.
In this process, the characteristics of the 2D electron dynamics at the diffraction screen can be imprinted to the generated harmonics, which provides abundant opportunities for manipulating light-matter interaction.
In this work, we study single-slit diffraction of a high-intensity spatiotemporal optical vortex (STOV) - a beam carrying transverse orbital angular momentum (OAM).
We demonstrate that due to the spatiotemporal structure of the driver, it induces differential electron oscillations on the screen,
which conveys the transverse OAM to the high-order harmonic STOV beams. As a result, the topological charges of the harmonic beams are $l = nl_0$, where $l_0$ is the topological charge of the fundamental driving light, and $n$ is the harmonic order.
In addition, we show that by controlling the slit angle with respect to the driver's transverse OAM, the STOV light can exert a torque on the plasma, thus  providing a way to manipulate the transverse OAM orientation of the generated harmonics.
\end{abstract}
\maketitle

Light pulses carrying orbital angular momenta (OAM) are known as optical vortices \cite{Allen1992,Yao2011,Bliokh20151},
where the energy flow of the electromagnetic field circulates around a local axis, giving rise to a null intensity core and a phase singularity.
Typically, the OAM vector is parallel to the laser's propagation axis, resulting in a helical wavefront characterized by phase winding in the spatial domain.
Recent studies have revealed the existence of spatiotemporal optical vortices (STOVs) \cite{Bliokh2012,Jhajj2016,Hancock2019,Bliokh2021,Wan2023}, referring to optical light possessing transverse OAM orthogonal to the propagation axis, with phase singularities in a plane coupling spatial and temporal profiles.
The interaction of such vortex light beams with matter has attracted increasing attention \cite{Zhang2016,Guo2023,Shi2024}, particularly on optical torques and angular momentum transport, as they play an important role in manipulating matter and controlling its dynamics \cite{Han2018,Xu2024,Wu2025}.

When the vortex light reaches relativistic intensities, namely when the electron quiver motion in the laser field approaches the speed of light, the emerging nonlinear effects in turn modify the optical phenomena in association with the plasma dynamics \cite{Mourou2006}.
This is of great interest as it not only sheds light on laser-matter interaction at relativistic intensities, but also leads to extreme optical torque enhancement.
A prominent research area is high-order harmonic generation (HHG) by vortex beams interacting with solid plasmas. In order to conserve total angular momenta, the spin and orbital angular momenta of the driver must be transferred to the harmonic beams, with the plasma as a nonlinear medium \cite{Zhang2015,Denoeud2017,Li2018,Zhang2022}. This provides fundamental insights into the spin-orbit and orbit-orbit angular momentum interactions of relativistic light.
Typically, HHG via laser-plasma interaction relies on the reflection of a relativistically-strong laser beam on a plasma foil. The electron oscillations on the reflecting surface produce harmonic beams due to the relativistic Doppler effect. This is known as the relativistic oscillating mirror mechanism \cite{Bulanov1994,Lichters1996,Baeva2006}.

Recently, it has become clear that the diffraction of light at relativistic intensities can also produce harmonic beams \cite{Yi2021,Duff2020,Bacon2022} through the so-called ``relativistic oscillating window" (ROW) mechanism \cite{Yi2021}.
The underlying physics can be attributed to the laser-driven electron oscillation on the periphery of a diffraction aperture.
Despite the similarities, a unique feature of the ROW is that the peripheral electron dynamics is intrinsically two-dimensional, which could carry information of angular momentum. The pattern (chirality) of the 2D electron motion at the diffraction screen can then be imprinted onto the harmonic beams, which can be harnessed to control the angular momentum coupling between the driver and harmonics \cite{Yi2021,Trines2024}. However, the studies so far are limited to the generation of optical vortices with OAM parallel to the beam axis.

\begin{figure*}[t]
	\centering
	\includegraphics[width=0.95\textwidth]{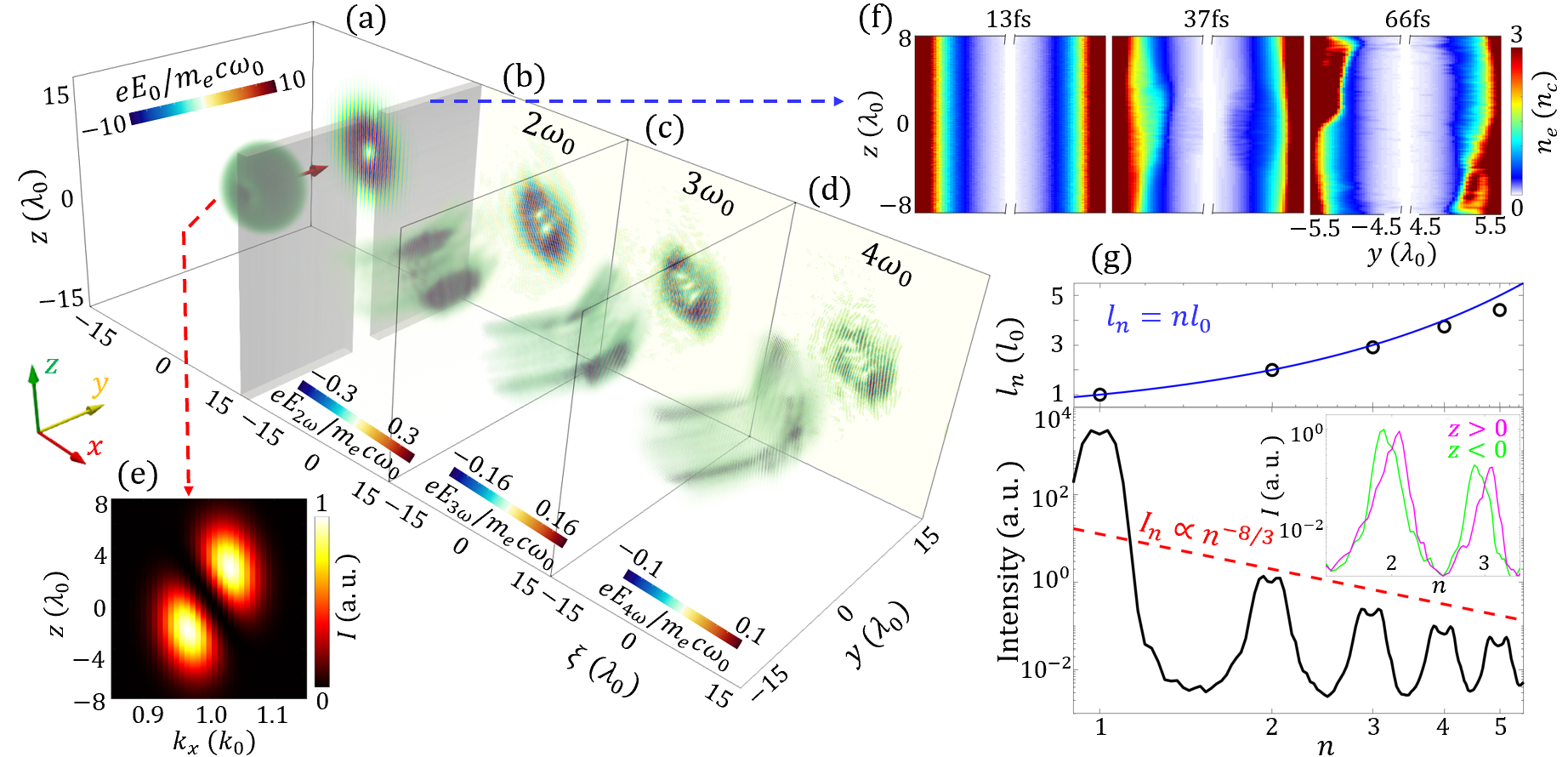}
	\caption{
		(a) Schematic setup of the STOV diffraction for slit angle $\alpha = 0$.
		The 3D color-coded donut represents the electric field distribution of the driving STOV beam, with its 2D cross section shown in the $\xi-z$ plane. The diffraction screen is shown in gray, and the red arrow indicates the transverse OAM of the driver.
		The second (b),third (c), and fourth (d) harmonic beams in the diffracted electromagnetic fields.
		(e) The 2D Fourier transformation showing the frequency distribution of the driver along $z$.
		(f) Snapshots of electron density on the diffraction screen at $t=13.3$, $37.0$, and $66.0\ {\rm fs}$, corresponding to the initial density distribution; when the rising edge of the driver arrives at the slit; and when the null point of the driver arrives at the slit, respectively.
%corresponding to the initial density distribution, and when the rising edge and the null point of the driver arrive at the slit, respectively.
		(g) Topological charge (upper) and the HHG spectra (bottom) of the diffracted electromagnetic fields. Inset: harmonic spectra of the diffracted fields in the $z>0$ and $z<0$ regions.
	}
\end{figure*}

In this letter, we consider the diffraction of a high-intensity STOV beam through a single slit. The driver could potentially be produced by tight focusing \cite{Chen2020,Rui2022} or laser interactions with solid targets \cite{Qiu2019,Chen2022}.
We show that when an STOV beam irradiates a diffraction screen, the laser-driven surface electron oscillations have different frequencies along the slit, forming a differential oscillating window that retains the spatiotemporal features of the drive pulse.
By means of 3D particle-in-cell (PIC) simulations, we demonstrate that such a differential oscillating window can convert the driver's transverse OAM to the harmonic beams,
and the orientation of the harmonic OAM can be controlled by the slit angle. This work offers a route towards full degree-of-freedom manipulation of light-matter interactions.

We first present our simulation setup and the results on HHG in Fig.~1. The PIC simulation is performed with the code EPOCH \cite{Arber2015}.
An STOV pulse traveling along the $x$ direction is focused onto the slit on a solid foil target.
The laser field is given as
\begin{equation}
\begin{aligned}
\mathbf{E}&=\mathbf{e_y}E_0 \left[ \sqrt{2\left(\frac{\xi^2}{w_\xi^2}+\frac{z^2}{w_z^2}\right)} \right] ^{\left| l \right|}\\
&\times
\exp\left[-\left(\frac{\xi^2}{w_\xi^2}+\frac{y^2}{w_y^2}+\frac{z^2}{w_z^2} \right)\right]
\exp\left( k_0\xi+l_0\varphi \right),
\end{aligned}
\end{equation}
where $\xi=x-ct$ is the propagation coordinate, $\varphi=\text{tan}^{-1}(z/\xi)$ is the spatiotemporal vortex phase, and here
we adopt a drive pulse with topological charge $l_0=1$.
%we use $l_0=1$ is the topological charge of the driver.
The STOV pulse is linearly polarized in the $y$ direction %($\mathbf{e_y}$)
(perpendicular to the slit).
$E_0$ is the laser amplitude, $\omega_0$ is the angular frequency, and $k_0 = 2\pi/\lambda_0$ is the wave number, with $\lambda_0 = 1\ \rm{\mu m}$ being the wavelength.
$\tau_0=\lambda_0/c$ is the laser period.
The spatial widths in all three directions are $w_{\xi}=w_y=w_z=5\ \rm{\mu m}$.
The laser has a normalized amplitude $a_0 = eE_0/mc\omega_0 = 16.5$,
where $c$ is the speed of light, $m$ is the electron mass, and e is the unit charge.

The single-slit diffraction screen is modeled by a preionized plasma target placed at $x_0=5.0\ {\rm \mu m}$, with electron density $n_0=30n_c$ for $|y|>y_0=6\ \rm{\mu m}$, where $n_c=\epsilon_0m\omega_0^2/e^2$ is the critical density. A density ramp at the inner boundary of the slit is modeled as
$n(y)=n_0\text{exp}[(|y|-y_0)/\sigma_0]$, with $\sigma_0=0.2\ \rm{\mu m}$ the scale length. The thickness of the screen is $0.5\ \rm{\mu m}$.
Here the slit angle, defined as the angle between driver's transverse OAM and the slit normal direction within the diffraction screen, is $\alpha = 0$ ($-\pi/2<\alpha\leq\pi/2$).
The simulation box has dimensions of $x\times y \times z = 30\times30\times30\ \mu m^3$ and is sampled by $4500\times600 \times600$ cells, with four macroparticles for electrons and two for C$^{6+}$ per cell.
A moving window is used to improve computational efficiency, which follows the propagation of the drive laser pulse.

\begin{figure*}[t]
	\centering
	\includegraphics[width=0.96\textwidth]{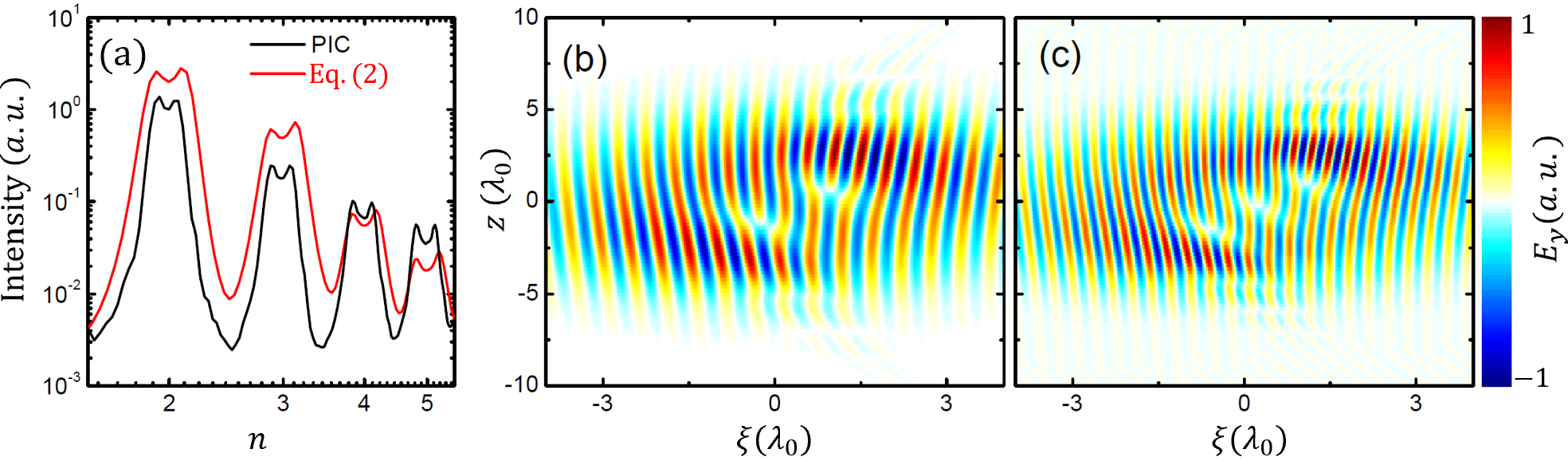}
	\caption{
		(a) A comparison of the harmonic spectra obtained from PIC simulation (Fig. 1) and the numerical calculations based on Eqs. (2-3).
		The 2D distribution of the second- (b), third-order (c) harmonic fields obtained from our numerical model, shown in the meridional plane ($\xi-z$).
	}
\end{figure*}

The 3D structure of the incident STOV pulse is depicted in Fig.~1(a), where one can see a donut-shaped pulse with one phase singularity at the center ($l_0 = 1$). The electromagnetic energy circulates around this singularity in the meridional plane, resulting in a transverse OAM (red arrow).
Previous studies \cite{Zhang2022,Guo2023} have shown that such a spatiotemporal profile arises because the upper-half ($z>0$) contains one extra optical cycle compared to the lower-half ($z<0$), corresponding to a slightly higher frequency shown by Fig.~1(e).

As a result, when such a STOV pulse travels through a single slit, it drives differential oscillation of the electrons on the rim, meaning the electrons are oscillating at different frequencies along the slit ($z$ direction), associated with the local driving pulse frequency.
This is demonstrated in Fig.~1(f), where the three panels (from left to right) show snapshots of electron density distribution on the screen at different times. A significant twist of the diffraction slit is observed due to frequency difference.
Such a differential oscillating window contains the information of the spatiotemporal phase structure of the driver, which can be imprinted onto the harmonic beams generated during the diffraction process. This will be discussed later.

Figures~1(b-d) illustrate the 3D electromagnetic field structures with harmonic order $n=2,3,4$, which are obtained by spectral filtering the diffracted electromagnetic wave that recorded at $x=40\ {\rm \mu m}$, in the frequency range $[n-0.5, n+0.5]\omega_0$.
Interestingly, all these harmonics carry transverse OAM, as indicated by the spatiotemporal holes in the $\xi-z$ plane owing to phase singularities. Moreover, the number of the holes, namely the topological charge is $l_n = nl_0$, which satisfies energy and angular momentum conservation.

The topological charge indicated by the spatiotemporal profile of the harmonic beams is cross-checked by the transverse OAM calculated from the retrieved complex-valued electromagnetic field \cite{Blinne2019,Bliokh20152}, as shown in the upper panel of Fig.~1(g). The lower panel shows the HHG spectrum from our PIC simulations,
which can be fitted with a power-law scaling $I_n\propto n^{-8/3}$
%This can be explained by nonlinear effects induced by ponderomotive force,
as expected from a linearly-polarized driver \cite{Yi2025}.
%In the reminder of this work, we focus on the normal diffracted harmonics, as the anomalously-polarized field intensity is one-order of magnitude smaller.
The HHG spectrum reveals a double-peak structure for each harmonic, this is also observed in the reflection of an STOV \cite{Zhang2022}.
As shown by the inset of Fig.~1(g), the higher and lower spectral peaks for each harmonic are associated with the electromagnetic wave recorded in the upper-half ($z>0$) and lower-half ($z<0$) domain, respectively. They correspond to the two frequency components above and below the phase singularity of the fundamental driver.\\

\begin{figure*}[t]
	\centering
	\includegraphics[width=0.96\textwidth]{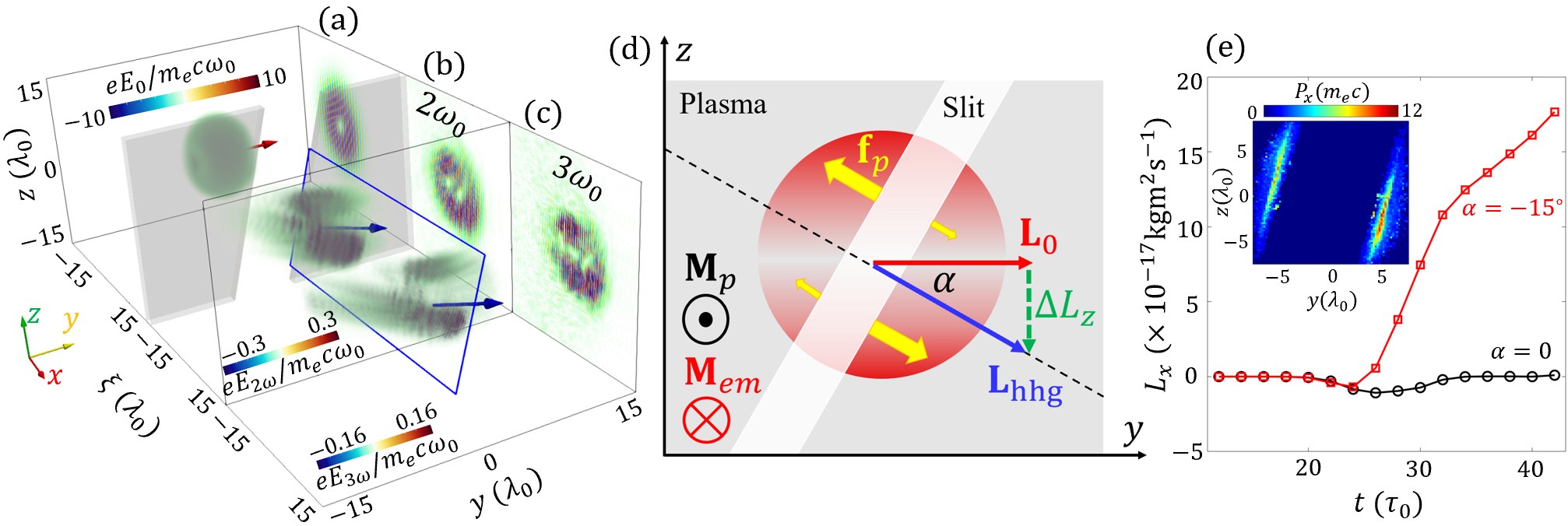}
	\caption{
		(a) Single-slit diffraction of an STOV beam for the slit angle $\alpha = -15^\circ$. The 3D color-coded electric field structure of the second (b) and third (c) harmonics, where the 2D cross-section view shows the field profile in the plane parallel to the single slit (represented by the blue box). The red and blue arrows in (a-c) represent the transverse OAM of the driver and the harmonics, respectively.
		(d) Schematic map showing the origin of the optical torque and rotation of electromagnetic angular momentum. The plasma is shown in gray, and the time-averaged STOV intensity is presented by the red (strong) and white (weak) colors. The yellow arrows show the time-averaged ponderomotive force acting on plasma, where the thickness represents the strength. The directions of the torques acting on the plasma and electromagnetic waves are represented by $\mathbf{M}_p$ and $\mathbf{M}_{em}$, respectively. The transverse OAM orientations of the driver and the harmonic beams are marked by $\mathbf{L}_0$ and $\mathbf{L}_{hhg}$, respectively, with $\Delta L_z$ representing the change in angular momentum along the $z$ direction.
		(e) Temporal evolution of the longitudinal angular momentum carried by the plasma during the STOV diffraction for $\alpha = 0$ and $-15^\circ$. The inset shows average longitudinal electron momentum distribution within the diffraction screen.}
		%the electric field profile of the second-order harmonic when $\alpha = 90^\circ$, which is no longer a STOV beam.}
\end{figure*}

In the following, we use the ROW model \cite{Yi2021} to explain the transverse OAM transport between the driver and the harmonics.
Taking into account the retarded effect induced by the oscillating window, the Kirchhoff integral of the diffracted harmonic fields can be written as \cite{Yi2025}
\begin{equation}
\mathbf{E_{hhg}}(\mathbf{r},t)=\frac{1}{2\pi}\nabla \times \int_{B}^{} \lbrack \mathbf{e_{n}} \times \mathbf{E}(\mathbf{r'},t')\rbrack \frac{{\rm exp}(ik_0R')}{R'} \, ds',
\end{equation}
where $(\mathbf{r}, t)$ and $(\mathbf{r}', t')$ represent the coordinates at the observer and the diffraction slit, respectively. $\mathbf{e_{n}}$ is the unit vector normal to the screen.
The distance between an area element at the slit $[ds'(x_0, y', z')]$ to an observation point $(x, y, z)$ on the screen is $R'(t')=|\mathbf{R}-d\mathbf{R'}(t')|$, where $\mathbf{R}$ is the initial distance, and $d\mathbf{R'}(t')$ denotes the displacement of $ds'$  measured at retarded time $t'=t-R'/c$ driven by the laser field.
Importantly, here the integration is not over an entire oscillating window that is assumed to be rigid \cite{Yi2021}, it is instead over a narrow bounded region near the rim of the slit ($\Delta y\sim0.5\lambda_0$). Such a quasi-1D area can be twisted to account for the deformation of the diffraction window induced by non-planar effects, which is crucial for the interaction in question.
We note that this method is justified as we are only interested in the harmonic beams generated via laser plasma interaction at the plasma-vacuum interface, thus the electromagnetic waves simply passing through in the middle are irrelevant.\cite{Yi2025}.

For the purpose of examining the spatiotemporal characteristics of the diffracted light, it is sufficient to apply the first order approximation that the surface electrons simply shift antiparallel to the driving laser field at the diffraction screen ($x=x_0$).
With the STOV driver defined in Eq.~(1), the displacement of $ds'$ can be expressed as
\begin{equation}
d\mathbf{R'}=- \mathbf{e_y} \delta_0 \exp[ik_0R'(t')-i\omega_0t+l_0\varphi'],
\end{equation}
where $\delta_0$ is the oscillation amplitude, and $\varphi'={\rm atan}[z'/(x_0-ct')]$.
Apparently, the spatiotemporal characteristics of the STOV driver are inherited by electron oscillation.
Since $\varphi'$ depends on $z'$ and $t$, it causes $d\mathbf{R'}$ to oscillate at different frequencies along the slit ($z'$).

Equation~(3) can be solved iteratively for $R'$ and substituted into Eq.~(2) to yield the diffracted field. We take $\delta_0 = 0.5\lambda_0$ for the surface wave-breaking case, and neglect nonlinear ponderomotive force for simplicity.
The results are presented in Fig.~2. The HHG spectrum calculated from our model [Fig.~2(a)] exhibits striking similarities to Fig.~1(g). Both show power-law spectral profiles, and present double-bump features at each harmonic peak.
Moreover, the numerical model successfully reproduces the exact number of the spatiotemporal holes on the harmonic field, and the cross-section plots of the harmonics [Figs.~2(b-c)] are in excellent agreement with Figs.~1(b-c).
This demonstrates that a diffraction window undergoing differential oscillation [Eq.~(3)] could convey its spatiotemporal characteristics to the transverse OAM of the harmonics.\\

%Finally, we show that adjusting the slit angle $\alpha$ manipulates the HHG process, exerting a significant optical torque on matter, and at the same time, allowing for control over the transverse OAM orientation of the harmonic beams.
Finally, we show that by adjusting the slit angle $\alpha$, the aforementioned HHG process can be manipulated to exert large optical torque on matter, and in the meantime, allows for control over transverse OAM orientation of the harmonic beams.
As illustrated in Fig.~3(a), we adopt a slit angle $\alpha = -15^\circ$, with other parameters the same as in Fig.~1. The 3D structures of the second and third order harmonics generated in this scenario are presented in Figs.~3(b-c), indicating that they both maintain the spatiotemporal structures of STOV beams with topological charges $l_2 = 2$ and $l_3 = 3$, respectively.
However, it is important to note that the 2D field distributions in Figs.~3(b-c) are plotted within a cross-section parallel to the slit (shown by the blue box), rather than in the initial meridional plane. This means the harmonic beams are rotated the same way as the slit, with their transverse momenta aligned with slit normal direction.

The underlying physical process for the rotation of harmonic STOV beams is sketched in Fig.~3(d).
Due to the null point of the drive STOV pulse, the intensity near the equator ($z=0$) is weaker, corresponding to a smaller time-averaged ponderomotive force ($\mathbf{f}_p$), as shown by the yellow arrows. When the slit is tilted, the asymmetry of the optical forces leads to an optical torque ($\mathbf{M}_p$) acting on the plasma, which results in the exchange of angular momentum between the plasma and the diffracted electromagnetic wave.

For the case presented in Fig.~3, $\mathbf{M}_p$ aligns with the $+x$ direction, thus the plasma rotates counter-clockwise. Consequently when the slit angle $\alpha\neq0$, the longitudinal angular momentum ($L_x$) of the plasma increases during the diffraction, as shown in Fig.~3(e).
On the other hand, the plasma must exert an opposite torque on the electromagnetic wave ($\mathbf{M}_{em}$) to ensure the conservation of total angular momentum.
This reaction explains the rotation of the harmonics observed in Fig.~3(b-c). However, since the STOV beams carry intrinsic transverse OAM, a rotation around the $x$-axis inevitably leads to a variation in the $z$-component of angular momentum $\Delta L_z$, as represented by the green dashed arrow in Fig.~3(d).
This is analogous to the rotation of a spinning wheel around an axis perpendicular to its spin.

The angular momentum variation $\Delta L_z$ indicates a corresponding torque acting on the electromagnetic wave,
which is only possible when the electrons originating from either side of the slit gain different longitudinal momenta $P_x$. 
Therefore, the diffracting electromagnetic wave receives a different ``kick" by these peripheral electrons during the HHG process.
This is confirmed by the 3D PIC simulations as shown in the inset of Fig.~3(e). With $P_x(y>0) > P_x(y<0)$, the electrons exert a torque in the $-z$ direction on the diffracting electromagnetic waves, which is responsible for the negative harmonic OAM along the $z$ direction. The physical reason for the difference in electron momenta lies in the fact that normalized laser amplitude $a_0$ is inversely proportional to the frequency. Thus due to slit rotation, the electrons originating at the lower-right side experience a higher $a_0$ than those at the upper left [Fig.~1(e)].\\

In conclusion, we have shown that high-order harmonics are generated when a relativistic STOV beam diffracts through a slit.
The spatiotemporal structure of the harmonic beams can be controlled by the slit angle $\alpha$.
When $\alpha=0$, the laser electric fields drive differential oscillation along the slit, which conveys the transverse OAM of the driver to the diffracted light, generating harmonic STOVs with topological charge $l_n = nl_0$.
When $\alpha\neq0$, the drive STOV beam exerts strong torque on the diffraction screen, making the plasma rotate around the beam axis.
In the meantime, the harmonic STOVs are rotated such that their transverse OAM are perpendicular to the slit.
This work sheds light on the fundamental nature of the STOV diffraction and the spatiotemporal coupling of relativistic laser-matter interaction. The proposed method paves the way for manipulating extreme optical torque on matter, and producing intense STOVs in the ultraviolet frequency range.

\begin{acknowledgments}
K. H. and X. G. contributed equally to this work. This work is supported by the National Key R$\&$D Program of China (No. 2021YFA1601700), and the National Natural Science Foundation of China (No. 12475246, 12205187).
\end{acknowledgments}

%\begin{acknowledgments}
%This work was supported by the Thousand Youth Talents Plan...
%\end{acknowledgments}

%\section*{aip publishing data sharing policy}
%The data that support the findings of this study are available from the corresponding author upon reasonable request.

\end{document}